\documentclass[aps,pre,twocolumn,showpacs,superscriptaddress,amsmath,amssymb]{revtex4}

\usepackage{graphicx}
\usepackage{color}
\usepackage{txfonts}
\usepackage{multirow,dcolumn}
\begin{document}

\title{Temporal network structures controlling disease spreading}

\author{Petter Holme}
\email{holme@skku.edu}
\affiliation{Department of Energy Science, Sungkyunkwan University, Suwon 440-746, Korea}

\begin{abstract}
We investigate disease spreading on eight empirical data sets of human contacts (mostly proximity networks recording who is close to whom, at what time). We compare three levels of representations of these data sets: temporal networks, static networks and a fully connected topology. We notice that the difference between the static and fully-connected networks---with respect to time to extinction and average outbreak size---is smaller than between the temporal and static topologies. This suggests that, for these data sets, temporal structures influence disease spreading more than static network structures. To explain the details in the differences between the representations, we use 32 network measures. This study concur that long-time temporal structures, like the turnover of nodes and links, are the most important for the spreading dynamics.
\end{abstract}


\maketitle

\section{Introduction}

The spread of infectious disease continues to be one of the major challenges to global health. Despite advances in genomics and access to large data sets, predicting outbreaks is still disturbingly difficult~\cite{difficult}. Even knowing fairly much about an outbreak (at the time of writing the major concern regards the zika virus~\cite{zika}) nobody can be very certain about its future. The basic mathematics of infectious disease outbreaks as emergent phenomena is well studied. No paper, to our knowledge, has strong alternatives to compartmental models---models dividing individuals into classes with respect to the disease and assigning transition rules between the classes. Straightforward implementations of compartmental models do, however, not explain the difficulties in predicting emergent outbreaks in real populations~\cite{jansson}. There can be many reasons for this difficulty to predict the extinction time outbreaks. One obvious reason is that the data quality is still not good enough to make high-precision forecasting. There can however be other, more fundamental issues with how the compartmental models are integrated with models of contact patterns (describing how people meet in such a way that disease can spread). In this paper, we investigate different levels of representing contact structures: as temporal networks (including information both of the time of contact and the individuals involved), as static networks (including informations of pairs of people between which the disease can spread), and fully-connected networks (which is the traditional contact structure of theoretical epidemiology~\cite{hethcote}).

Many studies have pointed out that to model disease spreading accurately, we need to understand both  static networks structures~\cite{keeling,vespi_rev} and temporal-network structures~\cite{masuda_holme_rev}. To make this point, a standard approach has been to first observe some structure in empirical data, then use models to prove this structure affects disease spreading, and finally conclude that this structure is important for epidemics. For example, Ref.~\cite{liljeros} observed power-law  distributions of degree (number of neighbors in the network) in sexual networks, Ref.~\cite{vespi:threshold} showed that model networks with power-law degree distributions need not to have an epidemic threshold, thus concluding the degree distribution is an important structure. For another example, Ref.~\cite{barabasi:burst} found power-laws in interevent time distributions, Ref.~\cite{small_slow,goh_vazquez} showed that outbreaks are slowed down by such fat-tailed distributions. Can we from this conclude that timing of contacts are important for disease spreading? Perhaps, but Ref.~\cite{holme_liljeros} argued that other, longer time-scale temporal structures are even more important. However, also Ref.~\cite{holme_liljeros} test two \textit{a priori} chosen structures. There could of course be other structures present affecting the spreading processes even stronger. The idea of this paper is to scan the possible structures in a less restrictive way, so as to open for the discovery of new important temporal network structures. For the same reason---that it is hard to \textit{a priori} reason about what the important temporal network structures are---we use empirical networks as our starting point rather than models generating the contact structure.

In this paper, we will run the Susceptible-Infectious-Recovered (SIR) compartmental disease spreading model (a canonical model for diseases that give immunity upon recovery) on eight human contact networks. We use temporal-network, static-network and fully-connected representations of these data sets. Then, to explain the deviations between the three representations, we explore 32 quantities measuring temporal-network structure.

\section{Preliminaries}

In this section, we will clarify the methods and precise model definitions in common to the rest of the paper. We also mention some computational considerations. In general, we assume a temporal network $H$ as input. It can be described as a list of $C$ \textit{contacts} $(i,j,t)$ where $i,j\in V$  are individuals and $t$ is the time of the contact (assuming a discretized time, as common for most data sets). The order of $i$ and $j$ does not matter. We set the smallest time to zero and label it $T$.

\subsection{SIR simulation}

We use the constant-duration version of the SIR model~\cite{holme:ploscomp}. In this model, a contact between a susceptible and infectious individuals infects the susceptible (instantaneously)  with a probability $\lambda$. Then the infectious  recovers a time $\delta$ later, and stays recovered for the rest of the simulation. The infection seed is chosen randomly and taken to become infectious immediately prior to its first contact. When there are no infectious individuals left, the infection is extinct. The time from the infection is introduced until the last recovery is the \textit{extinction time} $\tau$. The fraction of recovered individuals when the outbreak is extinct is the \textit{outbreak size} $\Omega$. 

For the static networks, we consider a disease spreading on a graph $G=(V,E)$ where $(i,j)\in E$ if there is a contact $(i,j,t)\in H$. We generate $C$ contacts between individuals connected by an link in $E$ at randomly chosen times between $0$ and $T$. Thus use as close as possible to the original data (assuming the maximum entropy principle---to maximize the randomness of the unknown structures). Analogously, for the fully-connected case, we also generate $C$ contacts at times in the interval $[0,T]$, but this time it can be between any pair of nodes.

Each data point of the $20\times 20$ parameter combinations is averaged over $200,000$ independent runs. We let the sequences of $\lambda$ and $\delta$ grow exponentially, as will be evident later, an exponential growth is needed to separate the data sets. For the same reason it is convenient to use the logarithm (we use the base-ten logarithm) of these values for discussion.

\subsection{Data sets}

\begin{table}
\caption{\label{tab:data}Basic statistics of the data sets. $N$ is the number of individuals; $C$ is the number of contacts; $T$ is the total sampling time; $t$-res.\ is the time resolution of the data set and $M$ is the number of links in the projected static networks.}
\begin{ruledtabular}
\begin{tabular}{l|rrrrr}
Data set & $N$ & $C$ & $T$ & $t$-res. & $M$ \\ \hline
\textit{Prostitution} & 16,730 & 50,632 & 6.00y & 1d & 39,044 \\
\textit{Conference} & 113 & 20,818 & 2.50d & 20s & 2,196 \\
\textit{Hospital}  & 75 & 32,424 & 96.5h & 20s & 1,139 \\
\textit{Reality}  & 64 & 26,260 & 8.63h & 5s & 722 \\
\textit{School 1} & 236 & 60,623 & 8.64h & 20s & 5,901 \\
\textit{School 2} & 238 & 65,150 & 8.58h & 20s & 5,541 \\
\textit{Gallery 1} & 200 & 5,943 & 7.80h & 20s & 714 \\
\textit{Gallery 2} & 204 & 6,709 & 8.05h & 20s & 739 \\
\end{tabular}
\end{ruledtabular}
\end{table}

As mentioned above, this study is based on empirical data sets of human proximity. In other words, they are recording two persons in close proximity at a certain time. For obvious reasons, these are interesting for disease spreading. We list the basic statistics---sizes, sampling durations, etc.---of the data sets in Table~\ref{tab:data}.

Our first data set (\textit{Prostitution}) comes from rom self-reported sexual contacts between female prostitutes and male sex buyers~\cite{prostitution}. This is a special form of proximity network in that a contact is sexual. Perhaps it should be classified as a separate type of network, but it is relevant for disease spreading. Several other data sets come from the Sociopatterns project (sociopatterns.org). These data sets by radio-frequency identification sensors that  record a contact when two sensors are within 1--1.5 m.  One of these datasets comes from  a conference~\cite{conference} (\textit{Conference}), another from a school (\textit{School})~\cite{school}, a third from a hospital (\textit{Hospital})~\cite{hospital} and a fourth from  an art gallery (\textit{Gallery})~\cite{gallery}. The \textit{Gallery} data set comprises 69 days where we use the first two. \textit{School} consists of two days and we use both.

A similar data set to the Sociopatterns data sets  comes from the Reality mining study~\cite{reality} (\textit{Reality}). Here contacts within a cohort of university students were recorded by the Bluetooth channel of smartphones. The range of such connections is between 10 and 15 meters. We use the same subset of data set as in Ref.~\cite{pfitzner}.

\subsection{Temporal network descriptors}

To characterize the temporal-network structures of the data sets, we use 32 different quantities, which we call network descriptors. We choose these both to be relatively simple and straightforward to interpret and to cover as wide spectrum of structures as possible. Table~\ref{tab:quantities} presents an overview of the descriptors.

\subsubsection{Time evolution}

We measure nine network descriptors characterizing the long-term behavior of the contact dynamics---briefly speaking, how the contact process differs from a stationary process. Some of these data sets (e.g.\ \textit{Prostitution}, \textit{Gallery 1} and \textit{2}) are growing throughout the sampling period, and this has been argued to  influence the spreading dynamics strongly~\cite{holme_liljeros}. In such a system, the disease could burn out in the population even before some individuals have entered it.

The first of these measures focuses on the time when nodes and links first appear in the data. First, we measure the fraction of nodes (links) present at half the sampling time relative to the final number of nodes, $f_{TN}$ (links, $f_{TL}$). Some studies argues the order of events is a more natural measure of time than the actual time. Thus we also measure the corresponding quantities $f_{CN}$ and $f_{CL}$ where half the sampling time is replaced by the half the contacts.

The second class of network descriptors, focuses on the persistence nodes or links. Let $F_{TN}$ ($F_{TL}$) be the fraction of nodes (links) present in the first and last $5\%$ of the time. The corresponding quantities for the sequence of contacts are $F_{CN}$ and $F_{CL}$.

Yet a measure related to the time evolution is the largest gap $g$ on the contact sequence. (During a gap, the disease cannot spread, and for long enough gaps, the disease could die out.)

\begin{table*}
\caption{\label{tab:quantities}Symbols and brief explanations of the network descriptors.}
\begin{ruledtabular}
\begin{tabular}{ll}
symbol & description \\ \hline
$f_{NC}$
& Fraction of nodes present (had been involved in least one contact) when half of the contacts happened. \\
$f_{NT}$ & Fraction of nodes present at half the sampling time.\\
$f_{LC}$ & Fraction of links present when half of the contacts happened.\\
$f_{LT}$ & Fraction of links present at half the sampling time.\\
$F_{NC}$ & Fraction of nodes present at both the first and last $5\%$ of the contacts.\\
$F_{NT}$ & Fraction of nodes present at both the first and last $5\%$ of the sampling time.\\
$F_{LC}$ & Fraction of links present at both the first and last $5\%$ of the contacts.\\
$F_{LT}$ & Fraction of links present at both the first and last $5\%$ of the sampling time.\\
$\mu_{Lt}$ & Mean link interevent time.\\
$\sigma_{Lt}$ & Standard deviation of interevent times of links.\\
$c_{Lt}$ &  Coefficient of variation of interevent times of links, also known as the average link burstiness~\cite{burstiness}. \\
$\gamma_{Lt}$ & Skewness of interevent times of links.\\
$\mu_{Ld}$ & Mean duration (time between first and last contact) of links.\\
$\sigma_{Ld}$ & Standard deviation of the duration of links. \\
$c_{Ld}$ & Coefficient variation of the duration of links. \\
$\gamma_{Ld}$ & Skewness of the duration distribution of links.\\
$\mu_{Nt}$ & Like $\mu_{Lt}$  but for nodes. \\
$\sigma_{Nt}$ & Like $\sigma_{Lt}$ but for nodes.\\
$c_{Nt}$ & Like $c_{Lt}$ but for nodes, i.e., the node burstiness. \\
$\gamma_{Nt}$ & Like $\gamma_{Lt}$ but for nodes. \\
$\mu_{Nd}$ & Like $\mu_{Ld}$  but for nodes. \\
$\sigma_{Nd}$ &  Like $\sigma_{Ld}$ but for nodes. \\
$c_{Nd}$ & Like $c_{Ld}$ but for nodes. \\
$\gamma_{Nd}$ & Like $\gamma_{Ld}$ but for nodes.\\
$g$ & The longest gap between any two contacts in the data.\\
$\mu_k$ & Average degree of the network of accumulated contacts. \\
$\sigma_k$ & Standard deviation of the degree distribution of the network of accumulated contacts.\\
$c_k$ &  Coefficient of variation of the degree distribution of the network of accumulated contacts. \\
$\gamma_k$ & Skewness of the degree distribution of the network of accumulated contacts.\\
$N$ & Number of nodes.\\
$C$ & Clustering coefficient of the network of accumulated contacts. \\
$r$ & Degree assortativity of the network of accumulated contacts.\\
\end{tabular}
\end{ruledtabular}
\end{table*}

\subsubsection{Node and link activity}

The node- and link-activity descriptors capture the bursty nature of human behavior. I.e.,  intense periods of activity separated by long periods of inactivity~\cite{burstiness}. One can imagine many ways to measure burstiness. The common starting point is interevent times---the time gap between consecutive contacts of a node or link. We measure four descriptors characterizing this kind of time series---the mean $\mu$, standard deviation $\sigma$, coefficient of variation $c$ (i.e.\ the standard deviation divided by the mean), and the  skewness
\begin{equation}\label{eq:skewness}
\gamma=\frac{(n^2-n)^{1/2}}{n-2}\frac{\mu_3}{\mu_2^{3/2}}
\end{equation}
where $\mu_2$ and $\mu_3$ are the second and third moment of the distribution, respectively.

Some studies have pointed out that the duration (time from the first to the last observation)  of nodes or links  can be important for spreading phenomena~\cite{holme_liljeros}. Therefore, we also study the distribution of node and link durations by the same four descriptors as the interevent times. In total, for this category, we define 16 network descriptors---$\mu$, $\sigma$, $c$ and $\gamma$, for both interevent-time and duration distributions and for both nodes and links.

\begin{figure*}
\includegraphics[width=0.9\textwidth]{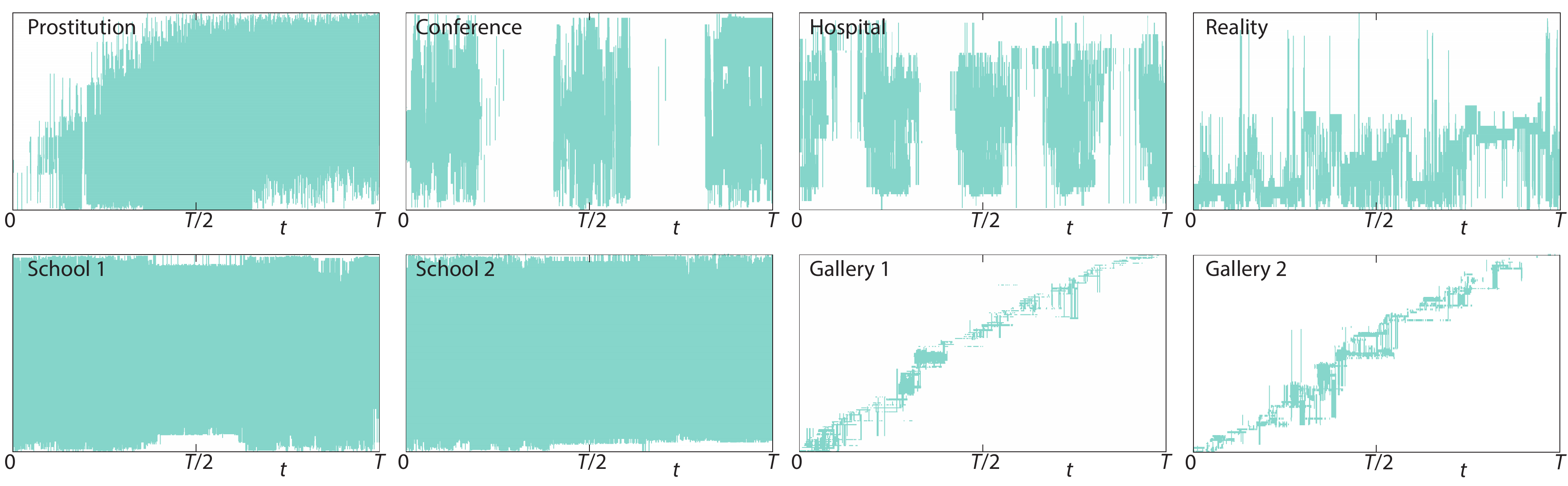}
\caption{(Color online) A visualization of the temporal structures of the data sets. The nodes are represented by a coordinate of the vertical axis. A contact at a certain time is displayed as a horizontal line between the coordinates of the nodes involved. The assignment of coordinates are optimized to reduce the total vertical distance of the  lines.}
\label{fig:timeline}
\end{figure*}

See Fig.~\ref{fig:timeline} for visualizations of the time structure of the data sets.

\subsubsection{Measures of static network structure}

How contact structures affect dynamic processes, such as epidemic spreading, is more established for static network structures than for temporal structures. We will measure the static network structure for the networks of accumulated contacts, i.e.\ if there has been at least one contact between two nodes we consider them connected by an link.

Arguably, the most important static network structure is the degree distribution describing how frequent it is to observe a node of a particular degree. Essentially a broad, right-skewed degree distribution (such as frequently observed in real systems) speeds up spreading phenomena~\cite{vespi_rev}. Usually researchers are interested in inferring the functional form of the degree distribution. For our purpose, we need to summarize the structures to numbers, no matter the functional forms. Therefore, we measure the same four quantities---$\mu$, $\sigma$, $c$ and $\gamma$---as for the interevent time and duration distributions.

In addition to the degree distribution, we also measure other static network descriptors. First, and simplest, the number of nodes, $N$. (But not the number of links since it is equal to $N\mu_{\rm deg}/2$. The next static network descriptor is the assortativity $r$. This is, in essence, the Pearson correlation of the degrees at either side of an link. One only has to symmetrize the arguments of the correlation coefficient (since the first and second arguments are different, but links are unordered with respect to the nodes---see Ref.~\cite{newman:book}) for details. The assortativity captures the tendency for nodes of similar degree to connect to each other. A large assortativity means that high-degree nodes connect to other high-degree nodes, and low-degree nodes to other low-degree nodes. It has been shown to have an influence on disease dynamics---assortative networks having lower epidemic thresholds~\cite{epi_ass}. Finally, we study the clustering coefficient---the number of triangles in the network normalized to the unit interval~\cite{newman:book}. A high clustering coefficient is known to slow down disease spreading~\cite{epi_clustering}.

\begin{figure*}
\includegraphics[width=0.9\textwidth]{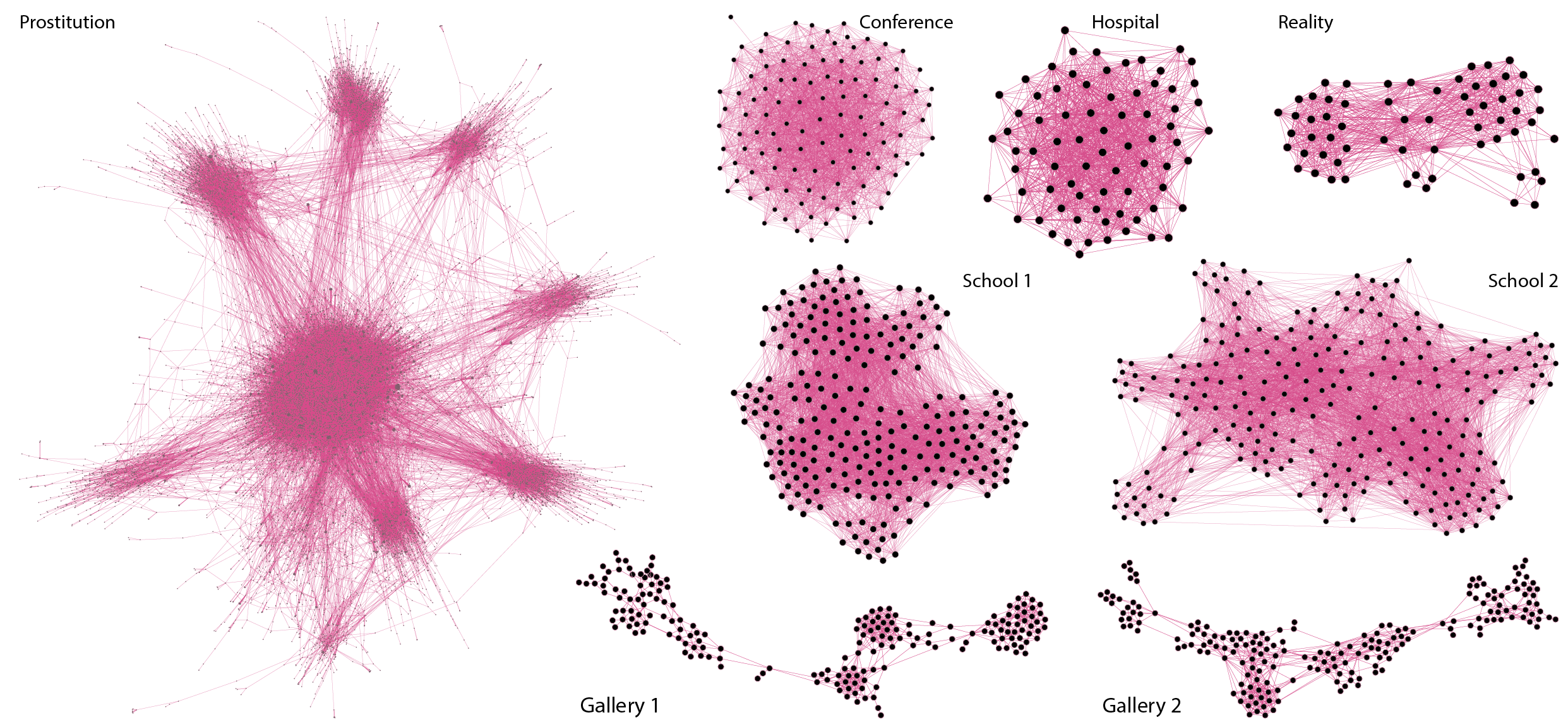}
\caption{(Color online) The network structure of the networks of aggregate contacts displayed using the ``Force Atlas 2'' method of the software package Gephi (gephi.org).}
\label{fig:structure}
\end{figure*}

See Fig.~\ref{fig:structure} for visualizations of the networks of accumulated contacts. Just like Fig.~\ref{fig:timeline}, this figure does not tell us more than that there are rich structures in the network topology that can influence the outbreak dynamics.

\subsection{Overlap statistics}

We will look at groups of data sets with different behavior of the SIR model with respect to the three levels of representations of the contacts. A good candidate network descriptor should separate the two groups well. With more samples, we could use e.g.\ the mutual information or Kullback-Leibler divergence, but with only eight data points, we can use a simpler quantity relying on the extreme values of the quantity for the two groups. Let $A$ be one subset of the data sets and $B$ its complement. Let $v(G)$ be the value for a quantity as a function of the data set $G$. Furthermore, assume (without loss of generality) that $\max_{G\in A}v(G)\geq\max_{G\in B}v(G)$ Then, more specifically, we measure
\begin{equation} \label{eq:x}
x_v(A,B)=\frac{\min_{G\in A}v(G)-\max_{G\in B}v(G)}{\max_{G\in A}v(G)-\min_{G\in A\cup B}} .
\end{equation}
In other words, if $\{v(G):G\in A\}$ and $\{v(G):G\in B\}$ do not overlap, then $x$ is the smallest difference between values in the two sets divided by the largest difference. If $x=1$, the separation is maximal. If $A$ and $B$ do overlap, $x$ will be negative, reaching a minimum $-1$ if the range of $\{v(G):G\in A\}$ and $\{v(G):G\in B\}$ are the same.

\section{Results}

\begin{figure*}
\includegraphics[width=0.9\textwidth]{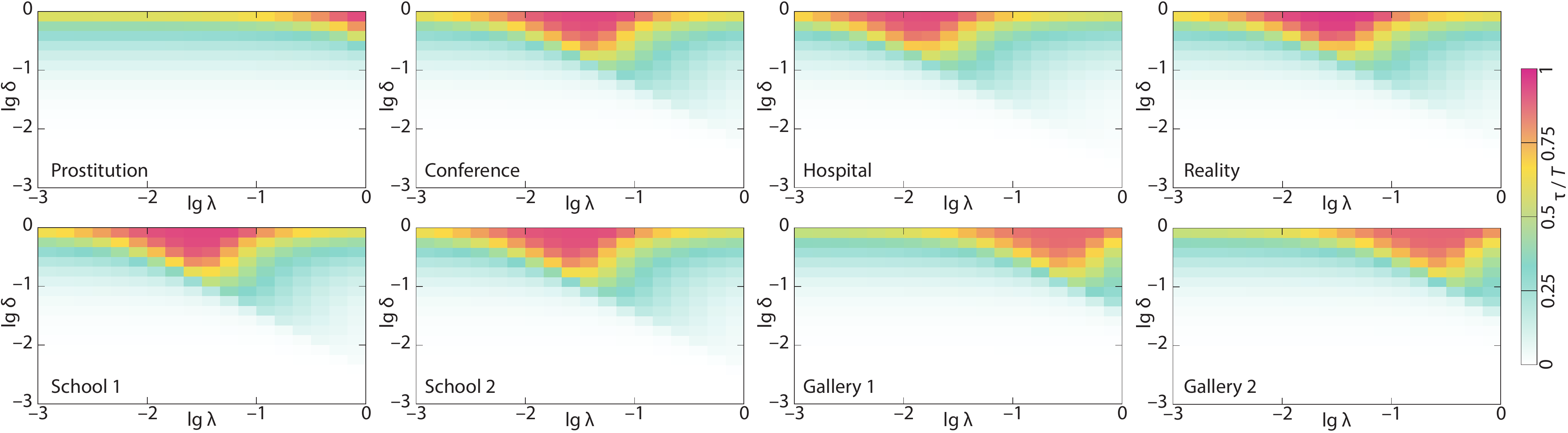}
\caption{(Color online) Average extinction times for the SIR model on eight temporal-network datasets of human proximity.}
\label{fig:tte_t}
\end{figure*}

\begin{figure*}
\includegraphics[width=0.9\textwidth]{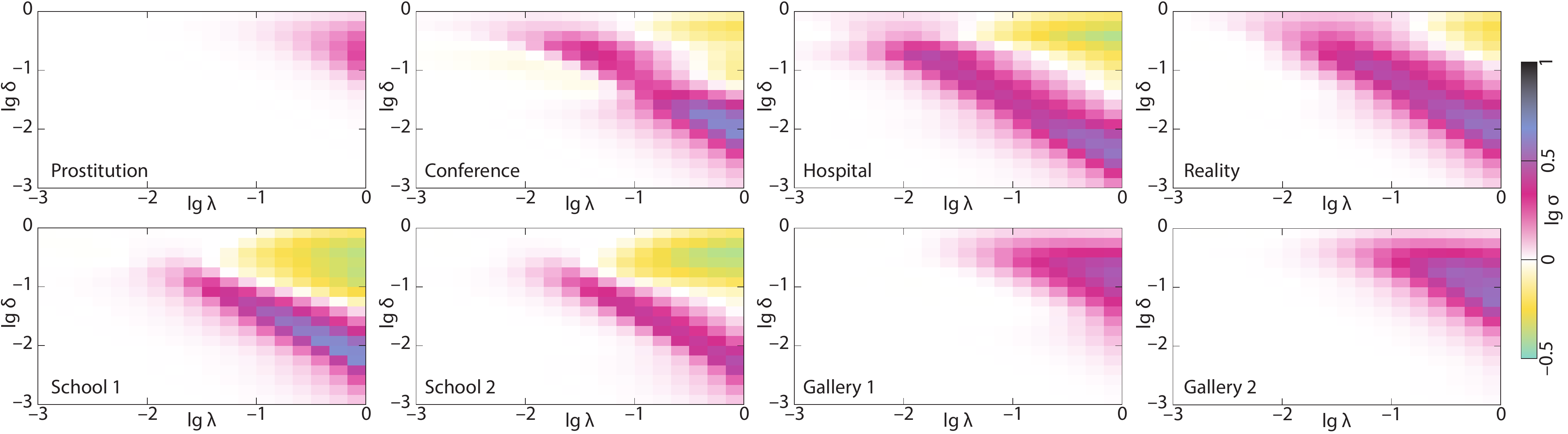}
\caption{(Color online) The difference between the average time to extinction of static and temporal network representations of the contact patterns.}
\label{fig:tte_diff_n}
\end{figure*}

\begin{figure*}
\includegraphics[width=0.9\textwidth]{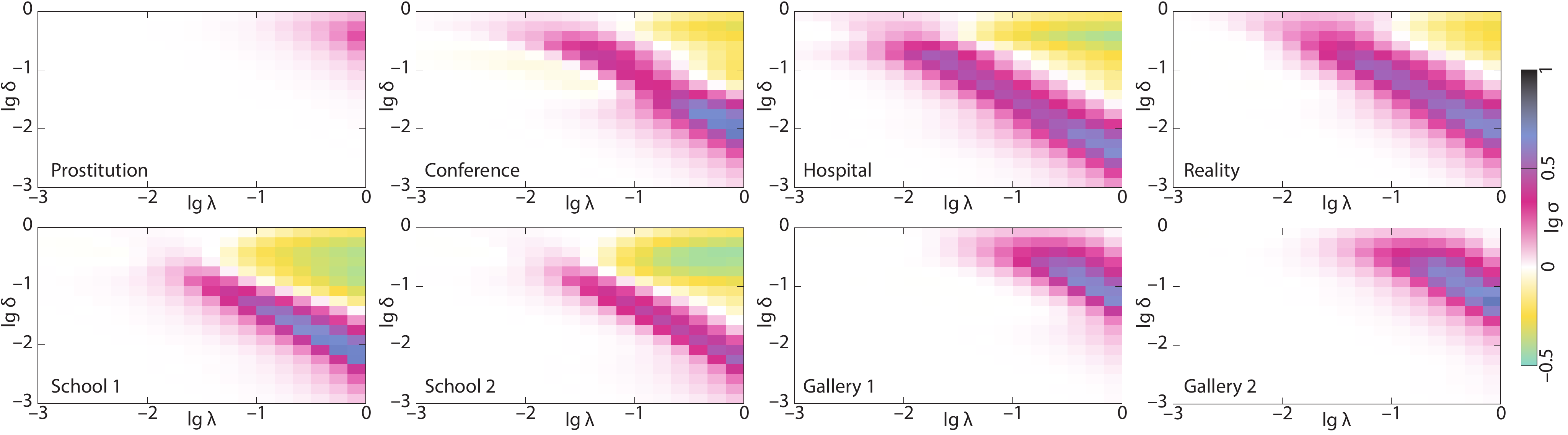}
\caption{(Color online) The difference between the average time to extinction of fully mixed and temporal-network representations of the contact patterns.}
\label{fig:tte_diff_fm}
\end{figure*}

\subsection{Extinction time}

One of our main quantities is the mean time to extinction $\tau$. Fig.~\ref{fig:tte_t} shows the values for SIR simulated on temporal-network representations. $\tau$ is strictly increasing with the disease duration $\delta$ but has a maximum in the per-contact transmission probability $\lambda$. The maximum comes from two conflicting mechanisms~\cite{holme_ext}. For small $\lambda$, decreasing $\lambda$ gives fewer chances for contagion and an increasing chance of the disease dying out. For large $\lambda$, the disease burns out fast in the population. The actual location of the peak varies much, from close to the maximum $\lambda=1$ for \textit{Prostitution} data set to $\lg\lambda\approx -1.8$ for the \textit{Hospital} data.

The effect of removing the temporal information by aggregating the contacts to a static network is seen in Fig.~\ref{fig:tte_diff_n}. This figure shows the deviation $\Delta\tau$ between $\tau$ of the static and temporal networks (so negative values means the outbreaks last longer in temporal networks). We see the \textit{Prostitution}, \textit{Gallery 1} and \textit{2} are different than the others in that they do not have regions of negative $\Delta\tau$---the static networks always give longer outbreaks. If we proceed, removing the network structure by making the network fully mixed (i.e.\ fully connected), then not much more happens (Fig.~\ref{fig:tte_diff_fm}). $\Delta\tau$ becomes larger for some regions of, in particular, the \textit{Gallery} data sets. The qualitative picture is, however, the same. Except for \textit{Prostitution}, \textit{Gallery 1} and \textit{2}, extinction times are underestimated for the largest $\lg\lambda$ and $\lg\delta$ and overestimated for intermediate $\lg\lambda$ and $\lg\delta$. This seems to suggest that the extinction time is more dependent on temporal than topological structures---a hypothesis we hope that future studies can confirm. Below, we will explore what separates the behavior of the \textit{Prostitution} and the two \textit{Gallery} data sets apart from the rest.

\begin{figure*}
\includegraphics[width=0.9\textwidth]{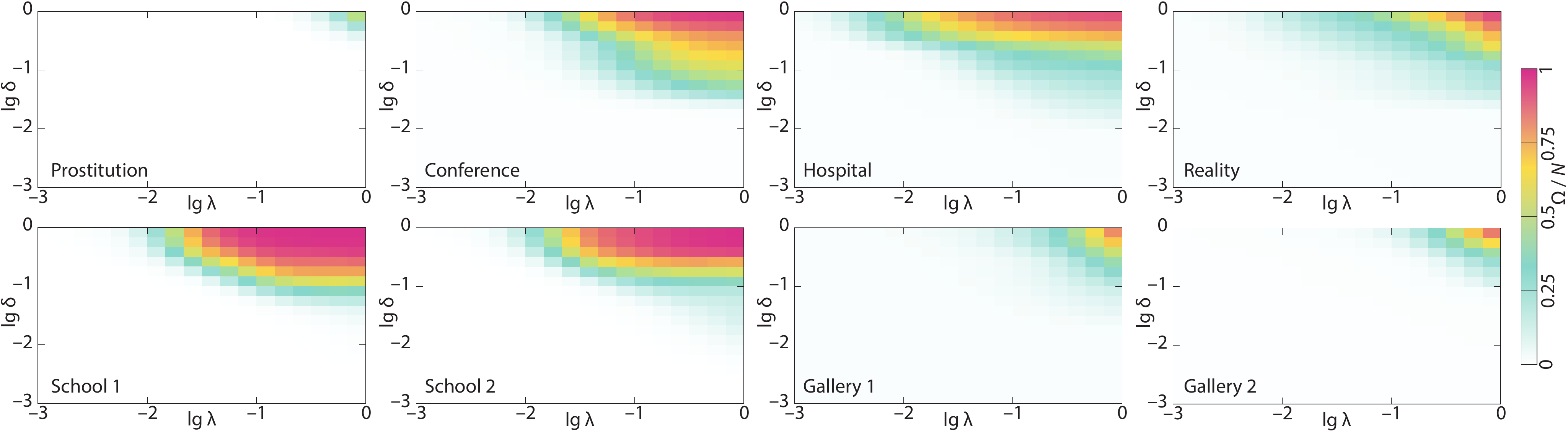}
\caption{(Color online) Average outbreak sizes for the SIR model on eight temporal-network datasets of human proximity.}
\label{fig:omega_t}
\end{figure*}

\begin{figure*}
\includegraphics[width=0.9\textwidth]{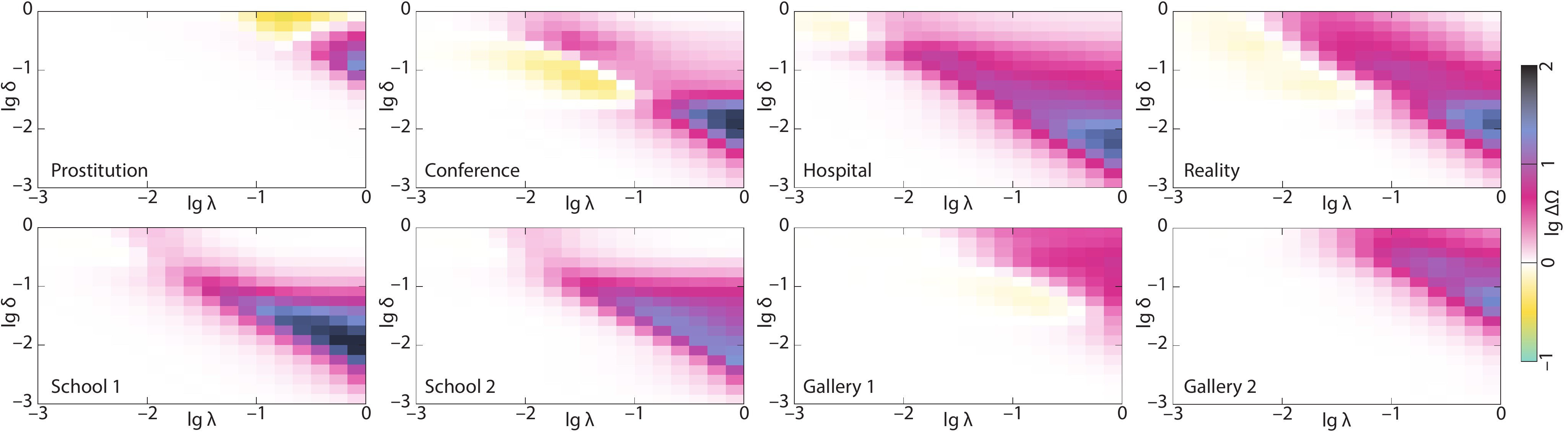}
\caption{(Color online) The difference between the average outbreak size of static and temporal network representations of the contact patterns.}
\label{fig:omega_diff_n}
\end{figure*}

\begin{figure*}
\includegraphics[width=0.9\textwidth]{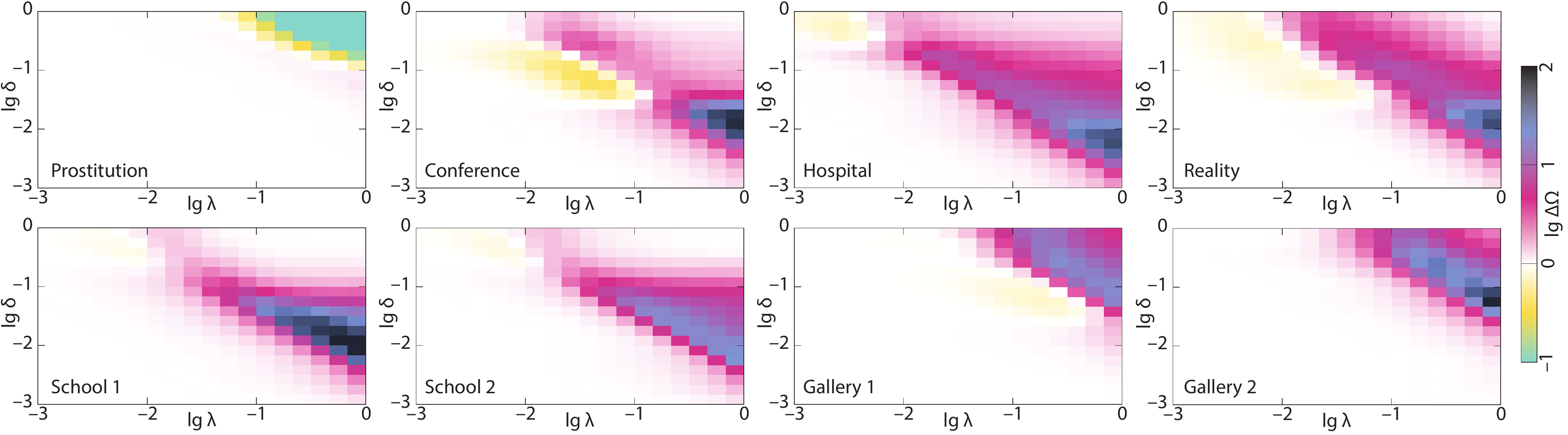}
\caption{(Color online) The difference between the average outbreak size of fully mixed and temporal-network representations of the contact patterns.}
\label{fig:omega_diff_fm}
\end{figure*}

\subsection{Outbreak size}

The average expected outbreak size $\Omega$ is perhaps a yet more common quantity than $\tau$ to characterize outbreaks in computational studies of disease spreading. Fig.~\ref{fig:omega_t} shows the values of $\Omega$ throughout the parameter space (for most data sets, these values were also presented in our Ref.~\cite{holme}). $\Omega$ is monotonically increasing with $\lg\lambda$ and $\lg\delta$ which probably is inevitable on average (even though, for specific seeds $i$, a larger $\lg\delta$ can lead to that the disease burn out so fast around $i$ that it is already extinct when  a contact leading away from $i$'s vicinity appears).

Figs.~\ref{fig:omega_diff_n} and \ref{fig:omega_diff_fm} show, respectively, the deviation when the temporal and both temporal and topological information is removed. Unlike $\tau$, the static network structure creates a qualitative difference---but only for the \textit{Prostitution} data. For this data set, the outbreak sizes are consistently underestimated for the fully connected networks, while for the static networks, $\lambda=\delta$ roughly separates two regions---for $\delta>\lambda$ the outbreak sizes are overestimated whereas for $\delta<\lambda$ they are underestimated. Below, we will look for a structural explanation behind this phenomenon.

\begin{figure}
\includegraphics[width=\columnwidth]{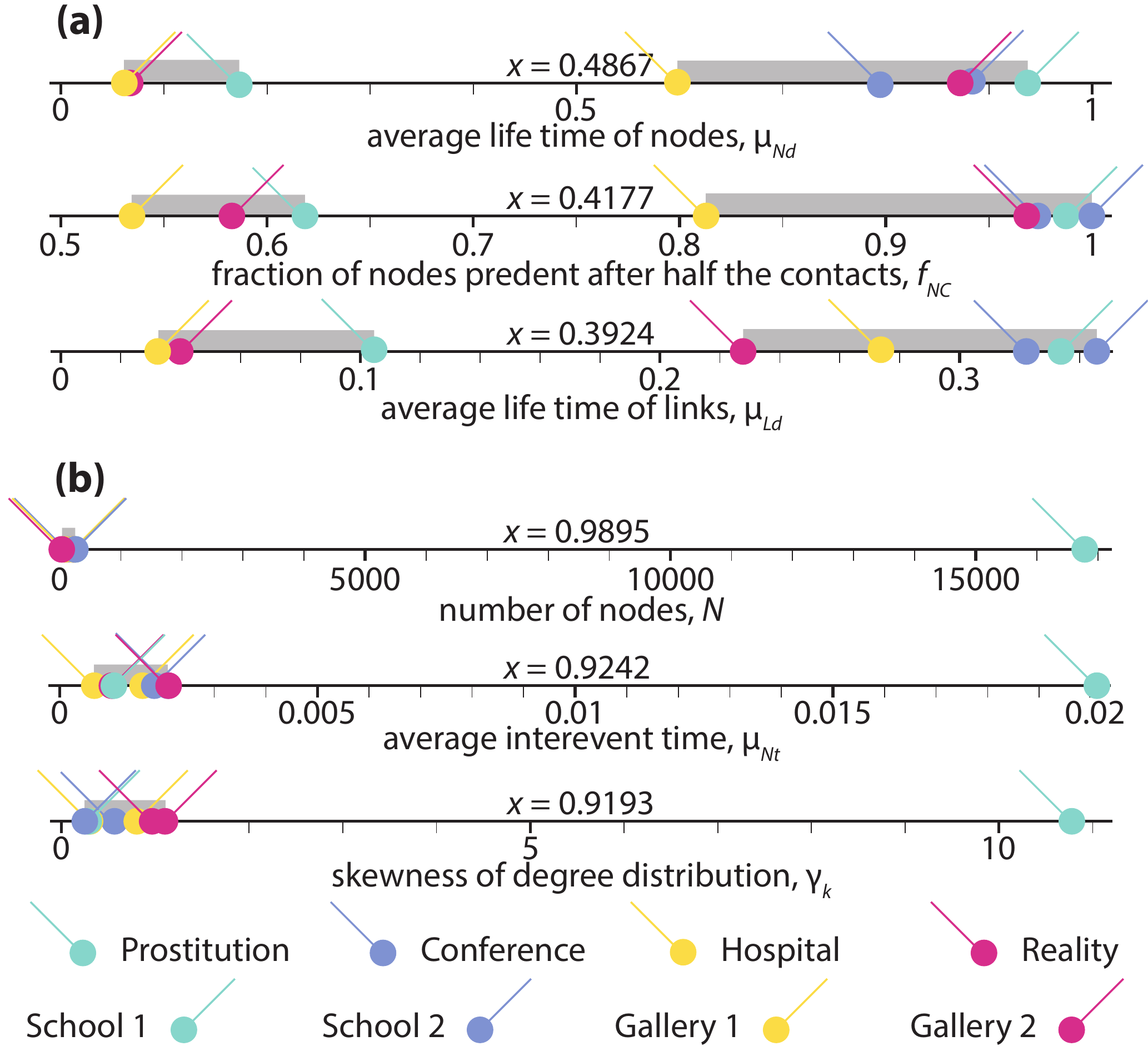}
\caption{(Color online) The top-three temporal-network structures separating the \textit{Prostitution}, \textit{Gallery 1} and \textit{Gallery 2} data sets from the rest of the data sets (a) and separating \textit{Prostitution} from the rest (b). $x$ is the difference between the smallest value (of the network structural measure in question) of the set containing the largest value, and the largest value of the other set divided by the difference between the largest and smallest values in the union of the sets.}
\label{fig:explanatory}
\end{figure}

\subsection{Structural explanations}

In this section, we we try to find what network structures that affect the effects found above. First, that \textit{Prostitution} and the \textit{Gallery} data sets differ from the others in that they lack a region of parameter space where the representations without temporal structure overestimates the time to extinction. Second, that \textit{Prostitution} has a different response to removing the network structure than all the other data sets. 

First, we investigate which network descriptors that separates $A=\{\mbox{\textit{Prostitution}}, \mbox{\textit{Gallery 1}}, \mbox{\textit{Gallery 2}}\}$ from the rest ($A$ refers to Eq.~\ref{eq:x} and the discussion about it).  The top three quantities $v$ with respect to $x_v$ along with their values, for the two groups of data sets are plotted in Fig.~\ref{fig:explanatory}(a).  These three quantities---the average life time of nodes $\mu_{Nt}$ and links $\mu_{Lt}$, and the fraction of nodes present at half of the contacts $f_{nC}$---are all temporal in nature, and all related to the turnover of individuals in the data, rather than higher frequency properties like the interevent time statistics. In more detail, we see that the data sets without regions of negative $\Delta\tau$ are characterized of a short average presence of the nodes and links in the data, and thus a high turnover of individuals. Representing such temporal networks as static networks destroys the long time-scale effects like that a node present early in the data cannot be infected by a node present only late in the data.

Our second investigation concerns how \textit{Prostitution} differs from the other data sets (Fig.~\ref{fig:explanatory}(b)). We find that the quantities with the largest $v$ values are: the number of nodes $N$, the average interevent time of nodes $\mu_{Nt}$ and the skewness of the degree distribution $\gamma_k$. These three quantities are very different from the ones to explain the other effect (in Fig.~\ref{fig:explanatory}(a)). The number of nodes is probably not an explanation for this effect in itself, but it could help accentuating other effects. The long average interevent times of \textit{Prostitution} come from a very skewed distribution of the number of contacts (a quantity we do not measure directly). The few-contact individuals can have long dormant periods, and thus increase the average interevent time. (Individuals with only one contact, of which there are around $35\%$, do not contribute to $\mu_{Nt}$.) The degree distribution is a very well-studied quantity, responsible for many peculiar features in static network epidemiology (such as the vanishing of epidemic thresholds or emergence of super spreaders~\cite{vespi_rev}). If is therefore reassuring to see its skewness as one of the top explanatory descriptors. It, furthermore, makes sense that the difference between the static networks and the fully connected networks is best explained by static network quantities. However, except the \textit{Prostitution} data, the static and fully connected-networks deviates from the temporal network in the same way, which means that the temporal structures are more influential with respect to disease spreading for these data sets, not only for $\tau$ but also for $\Omega$.

\section{Discussion}

We have compared SIR simulations (the entire parameter space) on three levels of representations of empirical contact data---temporal networks, static networks and fully-connected networks. We used two quantities characterizing the evolution of the outbreak---the time to extinction and the average outbreak size. We see that going from a temporal network representation to static- or fully connected network representations can lead to both a severe under- and over-estimation of both the extinction time and the average outbreak size. In general, short disease durations and high transmission probabilities lead to an over-estimation when the temporal information is discarded. Going from a static-network representation to a fully-connected topology does not make much of a difference  except for one data set (\textit{Prostitution}) and one of the quantities (average outbreak size). Looking closer at the quantities determining the patterns of over- and underestimation of $\tau$ and $\Omega$ also gives at hand that quantities describing the time evolution of the network are the most influential structures (in agreement to Ref.~\cite{holme}). Static network structure and shorter time scale temporal structure such as interevent times matters less. These observations are, of course, specific for the particular data sets we study. The results should be generalized with care. On the other hand, the contact data sets we use are as good as we can possibly obtain. There are no obvious structures in these data sets  that disqualify them as representative of real data sets (except, perhaps, the limited sizes). At the very least, this should encourage more research into the role of time structures in disease spreading.

There are many possible extensions of this work. Even though we used a generous amount of 32 network descriptors, one can imagine many other---describing how static network quantities change over the sampling time, how the activity level of nodes and their network position are correlated, etc. Ultimately, one would like to use results from this type of study to construct generative models for outbreak scenarios, retaining the important structures, but not more. Indeed, some such models have already been proposed~\cite{holme_saramaki,holme_modern}, but, to our knowledge, none that focuses on the longer time-scale features that we find important.

\begin{acknowledgments}
P.H. was supported by the Basic Science Research Program through the National Research Foundation of Korea (NRF) funded by the Ministry of Education (2013R1A1A2011947).
\end{acknowledgments}

\end{document}